\documentclass{jpsj-suppl}
\usepackage{graphicx}
\usepackage[dvips]{color} 

\title{\bfseries Charge-noise-free Lateral Quantum Dot Devices \\with Undoped Si/SiGe Wafer}

\author{T. Obata$^{1}$, K. Takeda$^{1}$, J. Kamioka$^{2}$, T. Kodera$^{2,3}$, W.M. Akhtar$^{1}$, \\ K. Sawano$^{4}$, S. Oda$^{2}$, Y. Shiraki$^{4}$, and S. Tarucha$^{1,5}$.}

\inst{$^{1}$Department of Applied Physics, The University of Tokyo, Hongo, Bunkyo-ku, Tokyo, 113-8656, Japan\\
$^{2}$Quantum Nanoelectronics Research Center, Tokyo Institute of Technology, O-okayama, Meguro-ku, Tokyo 152-8552, Japan\\
$^{3}$PRESTO, Japan Science and Technology Agency (JST), Honcho Kawaguchi, Saitama, Japan\\
$^{4}$Research Center for Silicon Nano-Science, Advanced Research Laboratories, Tokyo City University, 8-15-1 Todoroki, Setagaya-ku, Tokyo 158-0082, Japan\\
$^{5}$RIKEN, Center for Emergent Matter Science (CEMS), Wako, Saitama, 351-0198, Japan\\
}

\email{obata@meso.t.u-tokyo.ac.jp}

\date{today}

\abst{
We develop quantum dots in a single layered MOS structure using an undoped Si/SiGe wafer. 
By applying a positive bias on the surface gates, electrons are accumulated in the Si channel. 
Clear Coulomb diamond and double dot charge stability diagrams are measured. 
The temporal fluctuation of the current is traced, to which we apply the Fourier transform analysis. 
The power spectrum of the noise signal is inversely proportional to the frequency, and is different from the inversely quadratic behavior known for quantum dots made in doped wafers. 
Our results indicate that the source of charge noise for the doped wafers is related to the 2DEG dopant.
}

\kword{SiGe, Quantum dot, undoped wafer}

\begin{document}
\maketitle

\section{Introduction}
Single electron spin is one example of a two level system required to represent an elemental bit in quantum information processing. 
Important progress such as coherent manipulations of individual electron spins and elemental quantum gate operations are presented\cite{Petta,Koppens,MichelNP,ObataPRB}, which drives studies of lateral double quantum dot (DQD) with Si/SiGe\cite{SiGeSB,HRL} because one can make the decoherence time extremely long with isotopically purified Si/SiGe\cite{Wild} without the need for any nuclear spin feed-back techniques\cite{Rudner,Bluhm,ObataNJP}. 
We previously fabricated lateral DQDs in a modulation doped Si/SiGe wafer and observed charge noise, which is one of the central problems in doped materials\cite{Buizart,Takeda}. 
In order to further study the charge noise problem, we have fabricated lateral dot MOS devices using an undoped Si/SiGe wafer. 
 We characterize the noise property by adopting a frequency analysis of the current through DQDs and find the power spectrum is inversely proportional to the frequency, not inversely quadratic. 
 The latter is characteristic to the random telegraphic noise signals\cite{Buizart} therefore our results indicate the absence of telegraphic charge noise. 
 This finding can suggest that the low frequency or telegraphic charge noise arises from ionized donors, and offer a stable sample fabrication method using an undoped Si/SiGe wafer.

\begin{figure}
\begin{center}
\begin{minipage}{0.38\textwidth}
(a)
\includegraphics[width=0.9\textwidth]{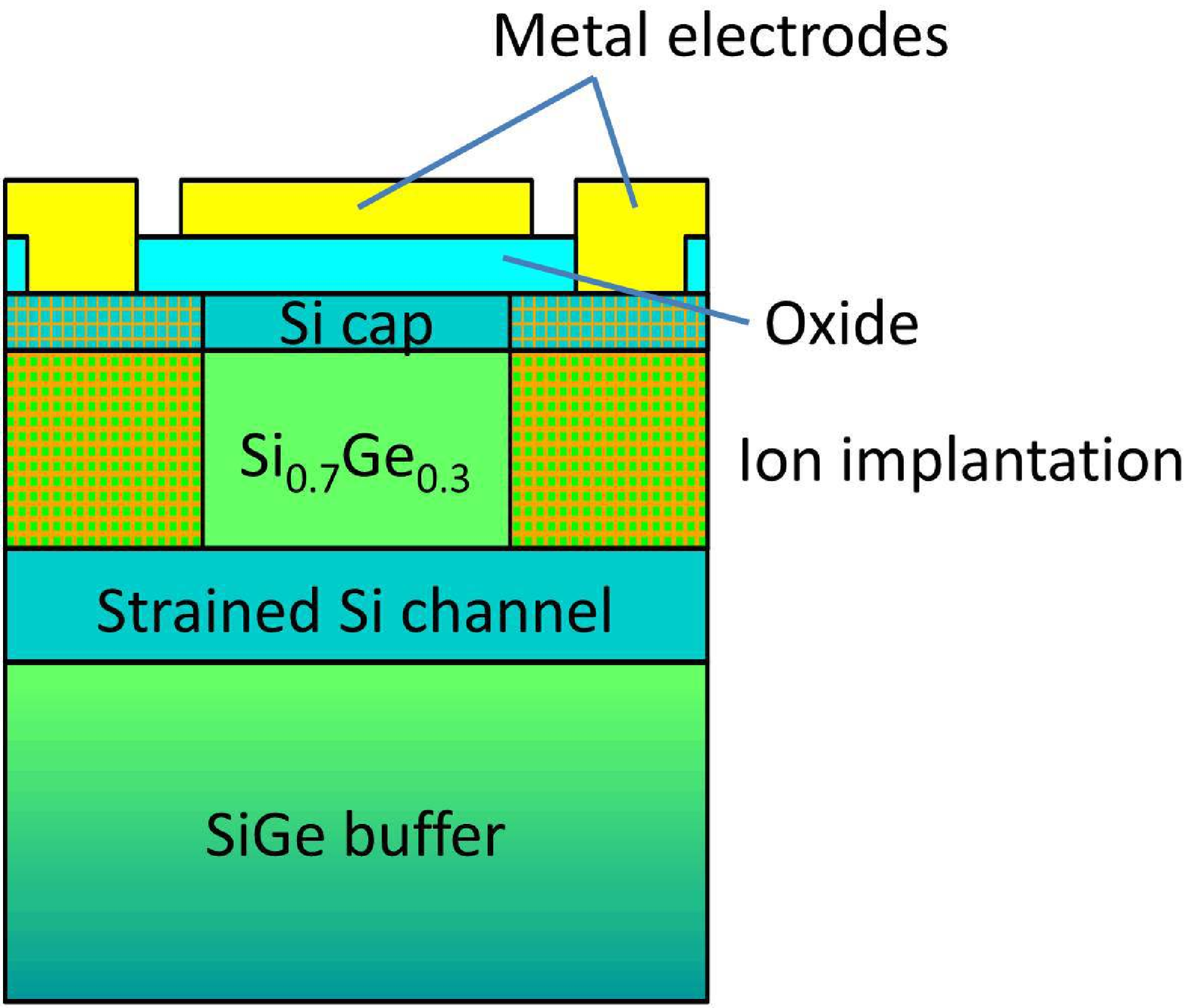}\\
\end{minipage}
\begin{minipage}{0.38\textwidth}(b)
\includegraphics[width=0.85\textwidth]{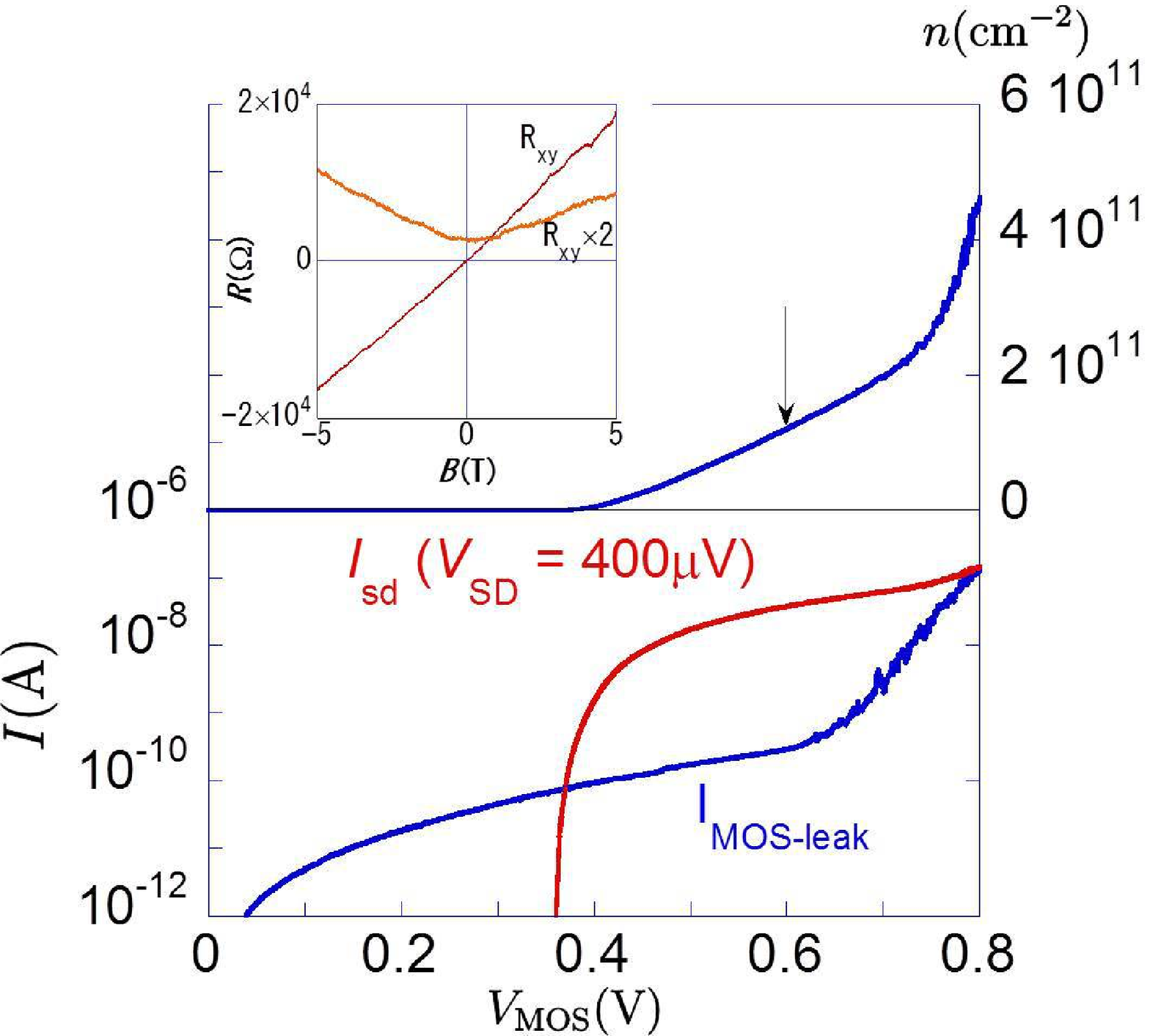}\\
\end{minipage}
\caption{Schematic drawing of the sample and a channel conductance.
(a) The sample has a MOS structure of a narrow channel in which a finite number of electrons are accumulated. (b) Gate bias dependence of the conductance through a MOS channel and calculated 2DEG density. The electron density is also calculated by Hall measurement (inset) at the position indicated by an arrow. }
\end{center}
\end{figure}

\section{Sample fabrication}

We use a CVD grown undoped wafer (Fig. 1a) whose layer stacks are; 1. Si substrate, 2. 3$\rm \mu m$ thick graded buffer layer in which the germanium concentration linearly increases from 0 to 30\%, 3. 1 $\rm \mu m$ thick Si$_{0.7}$Ge$_{0.3}$ buffer layer, 4. 15nm thick Si channel 5. 60nn thick Si$_{0.7}$Ge$_{0.3}$ without any dopant, 6. Si capping layer. 
The impurity concentration is well controlled to be less than $10^{14}$ atoms/cm$^{-3}$.

We micro-fabricate our sample on this undoped wafer. 
The active area for the 2DEG is restricted in a mesa area by reactive ion etching. 
Ohmic contacts are grown by ion-implantation of Antimony, followed by thermal annealing for re-allocation of the lattice. 
These contacts are covered with thin titanium and gold metals to protect the ohmic contacts against the following etching process for ohmic contacts. 
Hafnium oxide thin layer is added by an atomic layer deposition. 
This oxide is etched by RIE at the positions of the ohmic contacts. 
We deposit MOS gate metals for accumulating 2DEG layer and quantum dots. 
This MOS gate is stabilized by post-annealing at 300 degrees Celsius. 

We first fabricated a Hall bar to estimate the effective 2DEG density and its mobility at low temperature. 
The measurement result of the Hall bar is shown in Fig. 1b. 
An electron gas is accumulated with $V_{\rm MOS}>0.4$V. 
We measured both transverse and longitudinal resistance by four-terminal measurement at 1.9K. 
The longitudinal resistance shows a small but clear undulation of the conductance. 
The transverse resistance shows the Hall resistance with small kinks around the conductance of $2 e^2/h$. 
The reason why the kink is not just at $2e^2/h$ would be due to imperfection of our wafer. 
When we increase the voltage, we have an abrupt increment of a strong leakage current which presumably comes down from the MOS gate to the boundary of the ohmic contacts. 
The typical 2DEG density and the mobility at $V_{\rm MOS}=0.6\rm V$ is $1\times10^{11} \rm cm^{-2}$ and $1 \times 10^5 \rm cm^2/V\cdot s$.

\section{Coulomb oscillations}
\begin{figure}
\begin{center}
\begin{minipage}{0.40\textwidth}(a)
\includegraphics[width=\textwidth]{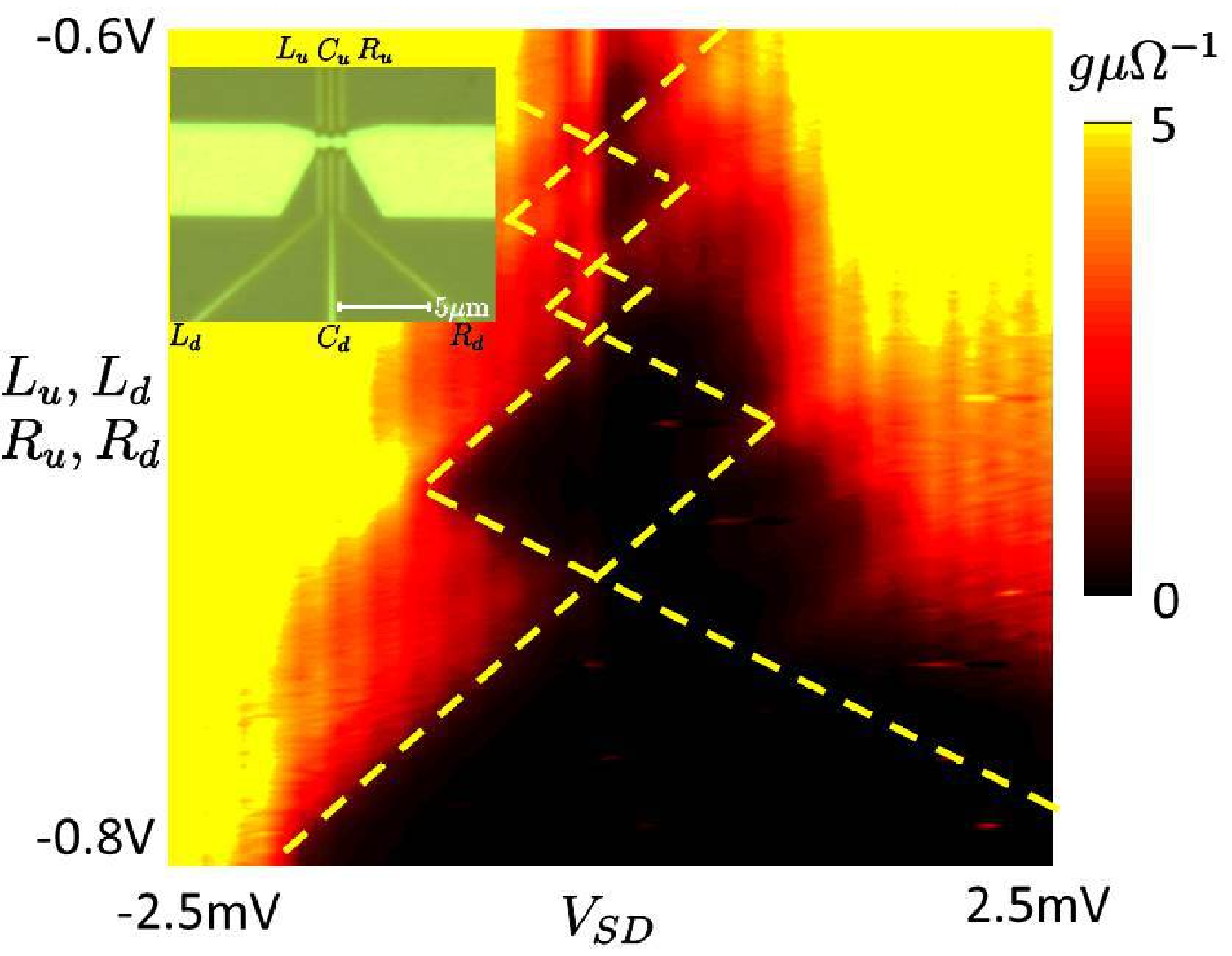}\\
\end{minipage}
\begin{minipage}{0.40\textwidth}(b)
\includegraphics[width=\textwidth]{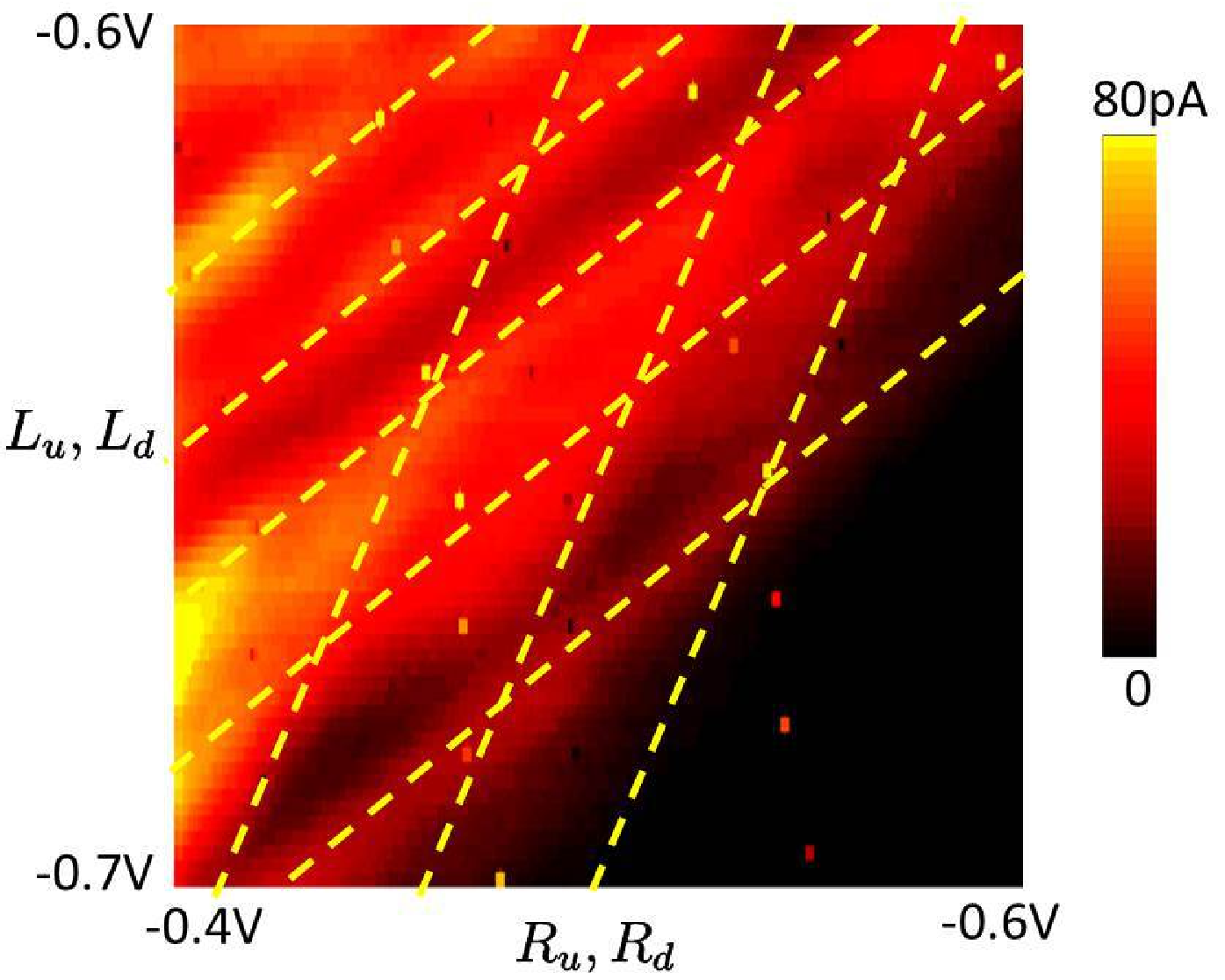}\\
\end{minipage}
\caption{ Coulomb oscillation and charge stability diagram.
(a) Source-drain spectroscopy of the device. The well-known Coulomb diamond structure is obtained. Large asymmetry of the diamond is detected. Inset: The micrograph of the device. The conductance is measured from left to right beneath the wide metal, while it is modulated by the finger gates $L_u,L_d,R_u,R_d,C_u,C_d$. (b) The conductance modulation by side gate scan. The honeycomb-like structure is measured. }
\end{center}
\end{figure}

We fabricate a MOS quantum dot device in Fig. 2(a), which has several surface metals. 
The large metal is used for accumulating the 2DEG while the side gates are used for squeezing the 2DEG and forming tunneling barriers. 
The 2DEG is accumulated by applying a sufficiently large bias on the global gate at 0.3K. 
By tuning the gate voltages, a quantum dot is found in Fig. 2(a). 
We measure the conductance. 
The source-drain spectroscopy reveals a Coulomb oscillation. 
The typical charging energy is about $\sim$1meV. 
An interesting feature found in Fig. 2 is the asymmetric shape of the Coulomb diamond and the asymmetric excited state around the side gate voltage of -0.7V. 
The asymmetric shape of the Coulomb diamond suggests the asymmetric shape of the electron orbital spread or the hypothesis that the quantum dot is a double quantum dot. 

In order to solve the internal mechanism of this asymmetry, we examined the side gate scan (Fig. 2(b)). 
We have observed two kinds of Coulomb peaks. 
One is modulated more by the voltages on $R_u$ and $R_d$ gates and the other is by $L_u$ and $L_d$. 
This indicates we have two quantum dots in series; {\it i.e.} one is closer to the left and the other is closer to the right. 
At the crossing points of these peaks we have an enhancement of the current, and the peak positions undulate slightly. 
This phenomenon is a strong indication of a coupled double quantum dot. 
The current at the crossing point is completely suppressed when we make the voltages on $C_u$ and $C_d$ less than -0.2V. 
The center barrier is carefully adjusted to observe the data of Fig. 2(b) by tuning the voltages on $C_u$ and $C_d$. 
In order to resolve the Coulomb peaks and the internal strtucture in the peaks due to the inter-dot coupling more clearly, we need to improve the gate structure to impose a strong confinement on the dots and a weak inter-dot tunnel coupling. 

\section{Frequency dependence of the current through the channel}
\begin{figure}
\begin{center}
\begin{minipage}{0.40\textwidth}
\includegraphics[width=\textwidth]{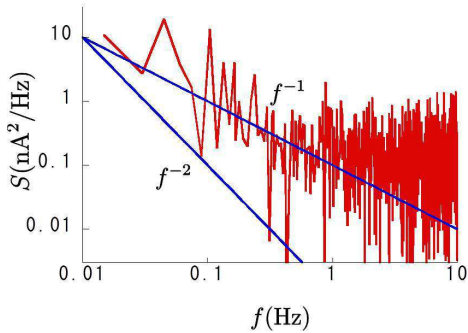}\\
\end{minipage}
\begin{minipage}{0.45\textwidth}
\caption{ Fourier analysis of the conductance through the channel. (a) Power spectrum of the signal at constant voltage on $V_{\rm MOS}$. The signal is measured at 4K. The conductance is set at 20k$\rm\Omega$ where the conductance is most sensitive to the fluctuation of MOS gate voltage. }
\end{minipage}
\end{center}
\end{figure}
As was expected, we haven't measured any clear charge noise behavior. 
In order to specify the noise characteristics in an undoped wafer, we get a temporal signal at a fixed conductance at 4K and apply the Fourier analysis. 
The power spectrum is shown in Fig. 3. 
The frequency dependence of $1/f$ behavior is found and differs from that of doped wafers \cite{Buizart,Takeda}. 
We applied finite bias on the side gates to squeeze the channel but we found no change of the dependence except for its magnitude. 
It is sometimes measured that the power spectrum of the signal shows $1/f^2$ behavior with doped wafers, which would have a two-level fluctuation of the conductance. 

We can calculate the voltage fluctuation in $V_{\rm MOS}$ by assuming all the conductance fluctuation is due to the gate fluctuation. 
The calculated value of this fluctuation is $\sim$ 10mV, which is 10 times larger than what we have reported\cite{Takeda}. 
We consider the reason that the MOS gate has much larger surface area to accumulate all the area for quantum dots and reservoirs. 
The larger the metal is, the larger the capacitance is formed between the metal and 2DEG, which lowers the impedance. 
This largely fluctuates the gate leakage current and so does the gate voltage. 
The same analysis is done on $V_{C_u}$ and $V_{C_d}$ and we find this contribution is negligibly  small, therefore the side gate voltage can be very stable. 
Our result instead suggests that the dopant layer in the doped wafer can attract a charge and temporarily disturb a constant flow of electrons in the tunneling process.

\section{Conclusion}

We have used undoped Si/SiGe wafers to fabricated quantum dots by a single layer MOS gate. 
A stable 2DEG is accumulated by the MOS gate and electrically confined by applying a finite bias on the side gates. 
The temporal fluctuation of the signal through the Si channel is analyzed by the Fourier analysis and the power spectrum shows no telegraphic noise behavior, which strongly indicates the doped material can produce both 2DEG and trapping sites for charge noise. 
Our finding can offer a method to stably produce a lateral quantum dot with Si/SiGe devices. 

\appendix
We gratefully thank J. Sailer, A. Wild, D. Bougeard, and G. Abstreiter for helpful discussions. 
This work was financially supported by GCOE for Physical Sciences Frontier, MEXT, Japan, Project for Developing Innovation Systems of the Ministry of Education, Culture, Sports, Science and Technology, MEXT, Japan, Grant-in-Aid for Scientific Research on Innovative Areas (21102003), MEXT, Japan, JSPS Grant-in-Aid for Young Scientists (B), Grant Number 24710148, and Funding for World-Leading Innovative R\&D on Science and Technology (FIRST) Program, Japan.


\begin{thebibliography}{9}
\bibitem{Petta} J. R. Petta, {\it et al.}, Science {\bf 309}, 2180 (2005). 
\bibitem{Koppens} F.H.L. Koppens {\it et al.}, Nature {\bf 442}, 766 (2006).  
\bibitem{Nowack} K.C. Nowack, F.H.L. Koppens, Y.V. Nazarov, and L.M.K. Vandersypen Science {\bf 318}, 1430 (2007).  
\bibitem{MichelNP} M. Pioro-Ladri{\`e}re {\it et al.}, Nat. Phys. {\bf 4}, 776 (2008). 
\bibitem{ObataPRB} T. Obata {\it et al.}, Phys. Rev. B {\bf 81}, 085317 (2010).  
\bibitem{Roland} R. Brunner {\it et al.}, Phys. Rev. Lett. {\bf 107}, 146801 (2011).  
\bibitem{SiGeSB} N. Shaji, {\it et al.}, Nat. Phys. {\bf 4}, 540 (2008).  
\bibitem{HRL} B. M. Maune, {\it et al.}, Nature {\bf 481}, 344 (2012).   
\bibitem{Buizart} C. Buizert, {\it et al.}, Phys. Rev. Lett. {\bf 101}, 226603 (2008).  
\bibitem{Takeda} K. Takeda {\it et al.},  Appl. Phys. Lett. {\bf 102}, 123113 (2013).  
\bibitem{Wild} A. Wild, {\it et al.}, Appl. Phys. Lett. {\bf 100}, 143110 (2012).  
\bibitem{Rudner} M.S. Rudner and L.S. Levitov Phys. Rev. Lett. {\bf 99}, 036602 (2007).  
\bibitem{Bluhm} H. Bluhm {\it et al.}, Nature Physics {\bf 7}, 109 (2010).  

\bibitem{ObataNJP} T. Obata, M. Pioro-Ladri{\`e}re, Y. Tokura and S. Tarucha New J. Phys. {\bf 14}, 123013 (2012).  
\end{thebibliography}
\end{document}